\documentclass[a4paper,aps,pra,showpacs,superscriptaddress,twocolumn]{revtex4-1}      
\usepackage[utf8]{inputenc}  
\usepackage[T1]{fontenc}     
\usepackage[british]{babel}  
\usepackage[sc,osf]{mathpazo}
\usepackage[scaled=0.86]{berasans}  
\usepackage[scaled=1.03]{inconsolata} 
\usepackage[usenames,dvipsnames]{color} 
\usepackage[colorlinks,citecolor=magenta,linkcolor=green,urlcolor=blue]{hyperref}  
\usepackage{graphicx} 
\usepackage[babel]{microtype}  
\usepackage{amsmath,amssymb,amsthm,bm,mathtools,amsfonts,mathrsfs,bbm} 
\usepackage{xspace}  
\usepackage{pgfplots}  
\DeclareMathOperator{\tr}{tr}
\DeclareMathOperator{\Tr}{Tr}

\newcommand{\bra}[1]{\left\langle #1 \right|}
\newcommand{\ket}[1]{\left| #1 \right\rangle}

\newcommand{\ketbra}[2]{\left|#1\middle\rangle\middle\langle#2\right|}

\newcommand{\ba}{\begin{eqnarray}}
\newcommand{\ea}{\end{eqnarray}}
\newcommand{\ban}{\begin{eqnarray*}}
\newcommand{\ean}{\end{eqnarray*}}

\begin{document}

\title{Quantum measurement incompatibility does not imply Bell nonlocality}

\author{Flavien Hirsch}
\affiliation{D\'epartement de Physique Th\'eorique, Universit\'e de Gen\`eve, 1211 Gen\`eve, Switzerland}
\author{Marco T\'ulio Quintino}
\affiliation{Department of Physics, Graduate School of Science, The University of Tokyo, Hongo 7-3-1, Bunkyo-ku, Tokyo 113-0033, Japan}
\author{Nicolas Brunner}
\affiliation{D\'epartement de Physique Th\'eorique, Universit\'e de Gen\`eve, 1211 Gen\`eve, Switzerland}

\date{\today}  

\begin{abstract}
We discuss the connection between the incompatibility of quantum measurements, as captured by the notion of joint measurability, and the violation of Bell inequalities. Specifically, we present explicitly a given a set of non jointly measurable POVMs $\mathcal{M}_A$ with the following property. Considering a bipartite Bell test where Alice uses $\mathcal{M}_A$, then for any possible shared entangled state $\rho$ and any set of (possibly infinitely many) POVMs $\mathcal{N}_B$ performed by Bob, the resulting statistics admits a local model, and can thus never violate any Bell inequality. This shows that quantum measurement incompatibility does not imply Bell nonlocality in general. 
\end{abstract}

\maketitle

The observation of quantum nonlocality---the violation of a Bell inequality by performing local measurements on an entangled state---has deep implications \cite{bell64,brunner_review}. In particular, for the bipartite case, Bell inequality violation implies that (i) the underlying quantum state must be entangled, (ii) the sets of local quantum measurements performed by one party (Alice) must be incompatible, and (iii) the sets of local quantum measurements performed by the other party (Bob) must be incompatible.

It is natural to ask whether the above links can be reverted. This can be formalized by two specific questions. The first one is whether entanglement implies Bell nonlocality, that is, whether for any entangled state one can find suitable sets of local measurements in order to violate some Bell inequality. This question, first discussed by Werner \cite{werner89}, has received a lot of attention \cite{brunner_review,augusiak_review}. The main result is the existence of entangled states admitting a local hidden variable (LHV) model for any possible local measurements \cite{werner89,barrett02}. This proves that entanglement does not imply Bell nonlocality. It should be noted that this statement concerns the scenario in which a single copy of the entangled state is used in every round of the Bell test; for more sophisticated scenarios (exploiting e.g. sequential measurements \cite{popescu95,hirsch16}, or many copies \cite{palazuelos12,cavalcanti12}), the relation between entanglement and nonlocality is still not established.

The second question focuses on the link between Bell nonlocality and quantum measurement incompatibility. Specifically, the question is whether any set of incompatible quantum measurements can be used in order to obtain Bell inequality violation \cite{Tsirelson85,anderson05,son05,wolf09}. Here one considers the most general class of quantum measurements, namely positive-operator-value-measures (POVMs). The incompatibility of a set of POVMs is characterized via the notion of joint measurability \cite{krausbook,buschbook,teiko08}. Loosely speaking, a set of POVMs $\mathcal{M}$ is said to be jointly measurable if there exists another POVM (called the mother POVM) such that all POVMs from the set $\mathcal{M}$ can be recovered as marginal (i.e. coarse-graining) of the mother POVM. 

The question can be precisely formalized as follows: given a set of non jointly measurable POVMs $\mathcal{M}_A$ performed by Alice, can one always find a shared entangled state $\rho$ and a set of (possibly infinitely many) POVMs $\mathcal{M}_B$ for Bob such that the resulting statistics violates a Bell inequality. In the present work, we show that the answer is negative, by constructing an explicit example. This shows that quantum measurement incompatibility (in the sense of non joint measurability) does not imply Bell nonlocality. 

This result has a number of implications. First, it shows that the non-classicality of quantum measurements is not sufficient in general to generate non-classical correlations, in the sense of violating a Bell inequality.  This is in contrast with recent results \cite{quintino14,uola14,uola15} proving a one-to-one relation between non joint measurability and quantum steering, a weaker notion of quantum nonlocality. Second, our work shows that the direct connection between the quantum violation of the CHSH inequality (the simplest Bell inequality) and non joint measurability of two binary POVMs proven in Ref. \cite{wolf09} does not hold in general. Finally, our work provides an example of a set of incompatible POVMs admitting a general LHV model. This generalizes previous work \cite{quintino15b} where such a model was constructed under the restriction of projective measurements for Bob.

\section{Preliminaries}

We first introduce notations. A set of $N$ POVMs, given by operators $M_{a|x}$ satisfying 
\begin{align} 
\sum_a M_{a|x} = \openone \, , \,\,\,  M_{a|x}\geq 0 \quad   \forall x \in \{1, \ldots , N \}
\end{align} 
is said to be jointly measurable if there exists one common POVM, $M_{\vec{a}}$, with outcomes $\vec{a}=[a_{x=1},a_{x=2},\ldots,a_{x=N}]$ where $a_{x}$ gives the outcome of measurement $x$, that is
\begin{align}
         M_{\vec{a}} \geq  0 , \quad \sum_{\vec{a}} M_{\vec{a}} =  \openone , \quad
        \sum_{\vec{a}\setminus a_x}M_{\vec{a}} = M_{a\vert x}  \;,
\end{align}
where the sum over $\vec{a}\setminus a_x$ means the sum over all elements of $\vec{a}$ except for $a_x$. The above equation implies that all $N$ POVMs $M_{a|x}$ are recovered as marginals of the \textit{mother POVM} $M_{\vec{a}}$. Notably, joint measurability of a set of POVMs does not imply that they commute \cite{kru}. 

Any set of POVMs for which no mother POVM can be defined is called incompatible, in the sense of being not jointly measurable. The incompatibility of a given set of POVMs can be characterized by semi-definite programming techniques (SDP) \cite{wolf09,paul_review}. Moreover, partial joint measurability does not imply full joint measurability in general \cite{krausbook}, contrary to commutation; see Refs \cite{teiko08,liang11} for elegant examples. More generally, it is known that any partial compatibility configuration can be realized in quantum theory \cite{fritz14}.

Measurement incompatibility plays a central role in quantum nonlocality. Consider two distant observers, Alice and Bob, performing local measurements on a shared entangled state $\rho$. Denote by $\mathcal{M}_A = \{M_{a|x}\}$ the set of POVMs performed by Alice, and similarly for Bob $\mathcal{N}_B = \{N_{b|y}\}$. 
The resulting statistics is given by 
\begin{equation}
p(ab|xy) = \tr(\rho M_{a|x} \otimes N_{b|y}). 
\end{equation}

It is straightforward to show that if the set $\mathcal{M}_A$ (or $\mathcal{N}_B$) is jointly measurable, then the resulting statistics necessarily admits a local hidden variable model \cite{quintino14,uola14}. That is, one can define a shared classical variable $\lambda$, distributed according to density $q(\lambda)$, and local response distributions 
$p_A(a\vert x,\lambda) $ and $ p_B(b \vert y,\lambda)$ such that

\begin{equation} \label{LHV}
    p(ab \vert xy)= \int d \lambda q(\lambda) p_A(a\vert x,\lambda) p_B(b \vert y,\lambda).
\end{equation}

In the present work we will show that the converse link does not hold in general.

\section{Main result} 

Our main result is to show explicitly a set of incompatible POVMs $\mathcal{M}_A$, such that for any quantum state $\rho$ and any set of (possibly infinitely many) POVMs $\mathcal{M}_B$ the resulting statistics admits a LHV model, i.e. a decomposition of the form \eqref{LHV}. 

Specifically, we consider the continuous set of dichotomic (binary) qubit POVMs, $\mathcal{M}_A^\eta  = \{ M_{\pm| \hat{x}}^\eta\}$, with elements
\ba \label{POVM}
M_{a| \vec{x}}^\eta = \frac{1}{2}  (  \openone  + a \eta \,  \hat{x} \cdot \vec{\sigma})
\ea
with binary outcome $a=\pm1$. Here $\hat{x} $ is any vector on the Bloch sphere denoting the measurement direction, and $\vec{\sigma} = (\sigma_1, \sigma_2, \sigma_3)$ is the vector of Pauli matrices. 

The set $\mathcal{M}_A^\eta$ features a parameter $0 \leq \eta \leq 1$, representing the level of noise of the measurements, or equivalently the purity of the POVM elements. For $\eta=1$, all POVM elements are projectors, while for $\eta=0$ the set contains only the maximally mixed POVM (both POVM elements being $ \openone/2$). More generally, it is known that the set $\mathcal{M}_A^\eta$ is jointly measurable if and only if $\eta\leq 1/2$ \cite{quintino14,uola14}. 

Here we prove that for $\eta^* \simeq 0.525$, the set $\mathcal{M}_A^{\eta^*}$ cannot lead to any Bell inequality violation. More precisely, we construct a LHV model considering any shared quantum state $\rho$ and arbitrary POVMs for Bob. Since $\mathcal{M}_A^{\eta^*}$ is not jointly measurable, this shows that measurement incompatibility does not imply Bell nonlocality.

\section{Proof}

The general structure of the proof is similar to that of Ref. \cite{quintino15b}. Nevertheless we describe here all steps for completeness. 

Our goal is to show that the statistics 
\ba \label{pQ}
p(ab|xy) = \tr(\rho M^{\eta^*}_{a|\hat{x}} \otimes N_{b|y})
\ea
is local for all possible quantum state $\rho$ and all possible POVMs $N_{b|y}$ on Bob's side. Here, Alice's POVM $M^{\eta^*}_{a|x}$ belongs to the set $\mathcal{M}_A^{\eta^*}$.

First, note that since the probabilities \eqref{pQ} are linear in $\rho$, and the set of local correlations is convex \cite{brunner_review}, one can focus on pure states. Also, given that $\mathcal{M}_A^{\eta^*}$ consists only of qubit measurements, Alice's subsystem can be considered to be a qubit. Then, since the set 
$\mathcal{M}_A^{\eta^*}$ is invariant under qubit rotations, we can freely choose the reference frame on Alice's side. Moreover, since we can also choose the reference frame on Bob's side, we can express the shared state $\rho$ in the Schmidt form, i.e. $\rho = \ket{\phi_\theta} \bra{\phi_\theta}$ with
\ba \label{pure_ent}
\ket{\phi_\theta} = \cos{\theta} \ket{00} +  \sin{\theta} \ket{11}
\ea
and $\theta \in [ 0 , \pi/4]$.

For the measurements of Bob, we need to consider all qubit POVMs $N_{b|y}$. Any such POVM can actually be viewed as a four-outcome qubit measurement followed by classical post-processing \cite{dariano05}. 

Therefore, our problem can be reformulated as follows. We must show that the statistics 
\ba \label{pF}
p(ab|xy) = \tr( \ket{\phi_\theta} \bra{\phi_\theta} M_{a|\hat{x}}^{\eta^*} \otimes N_{b|y})
\ea
is local for all $\theta \in [ 0 , \pi/4]$, all measurement directions $\hat{x}$, and all four-outcome qubit POVM $N_{b|y}$. From equation \eqref{POVM}, we have that:
\ba \label{map}
\tr( \ket{\phi_\theta} \bra{\phi_\theta} M_{a|\hat{x}}^{\eta^*} \otimes N_{b|y}) = 
\tr( \rho_\theta^{\eta^*} \Pi_{a|\hat{x}} \otimes N_{b|y})
\ea
where 
\ba \label{Maf}
\rho_\theta^{\eta^*} = \eta^* \ket{\phi_\theta} \bra{\phi_\theta} + (1 - \eta^* ) \frac{\openone}{2} \otimes \rho_B
\ea
and $\rho_B = \tr_A(\ket{\phi_\theta} \bra{\phi_\theta} )$. Note that we have now introduced projective qubit measurements $\Pi_{a|\hat{x}} = M^{\eta=1}_{a|\hat{x}}$. 

Thus, our problem has been mapped to the one of finding a LHV model for the class of state $\rho_\theta^{\eta^*}$. Importantly the LHV model must work for all $\theta \in [ 0 , \pi/4]$, for all projective measurements $\Pi_{a|\hat{x}}$ for Alice and for all 4-outcome POVMs $N_{b|y}$ for Bob.

It turns out that the local properties of the states of the form \eqref{Maf} have been recently discussed \cite{bowles15b}. However, in this work, local models could be constructed only for the case of local projective measurements. Note that this model is of the form of a local hidden state model, and can thus be straightforwardly extended to the case of POVMs for Alice and projective measurements for Bob.  Here, however, we must consider the opposite situation, where Alice performs projective measurements and Bob performs general POVMs. To construct such a model, we make use of the techniques of Refs \cite{hirsch15,cavalcanti15}, i.e. an algorithmic procedure for constructing local models for a given target entangled state. Specifically, we proceed in two steps. First, we will use the method to demonstrate that a given (finite) set of states $\rho_\theta^{\eta}$ (for specific values of $\theta$ and $\eta$) are local. This gives us a net of local states, well distributed over the interval $\theta \in [ 0 , \pi/4]$ (see Fig. 1). Second, to derive our final result, we show how the entire (continuous) interval $\theta \in [ 0 , \pi/4]$ can be covered using continuity arguments. This leads to the final result, namely that the states $\rho_\theta^{\eta^*}$ are local for $ \eta^* = 0.525 $ for all $\theta \in [ 0 , \pi/4]$. Below we discuss both steps in detail. 

\begin{figure}[b!] \begin{center}
\includegraphics[width=\columnwidth]{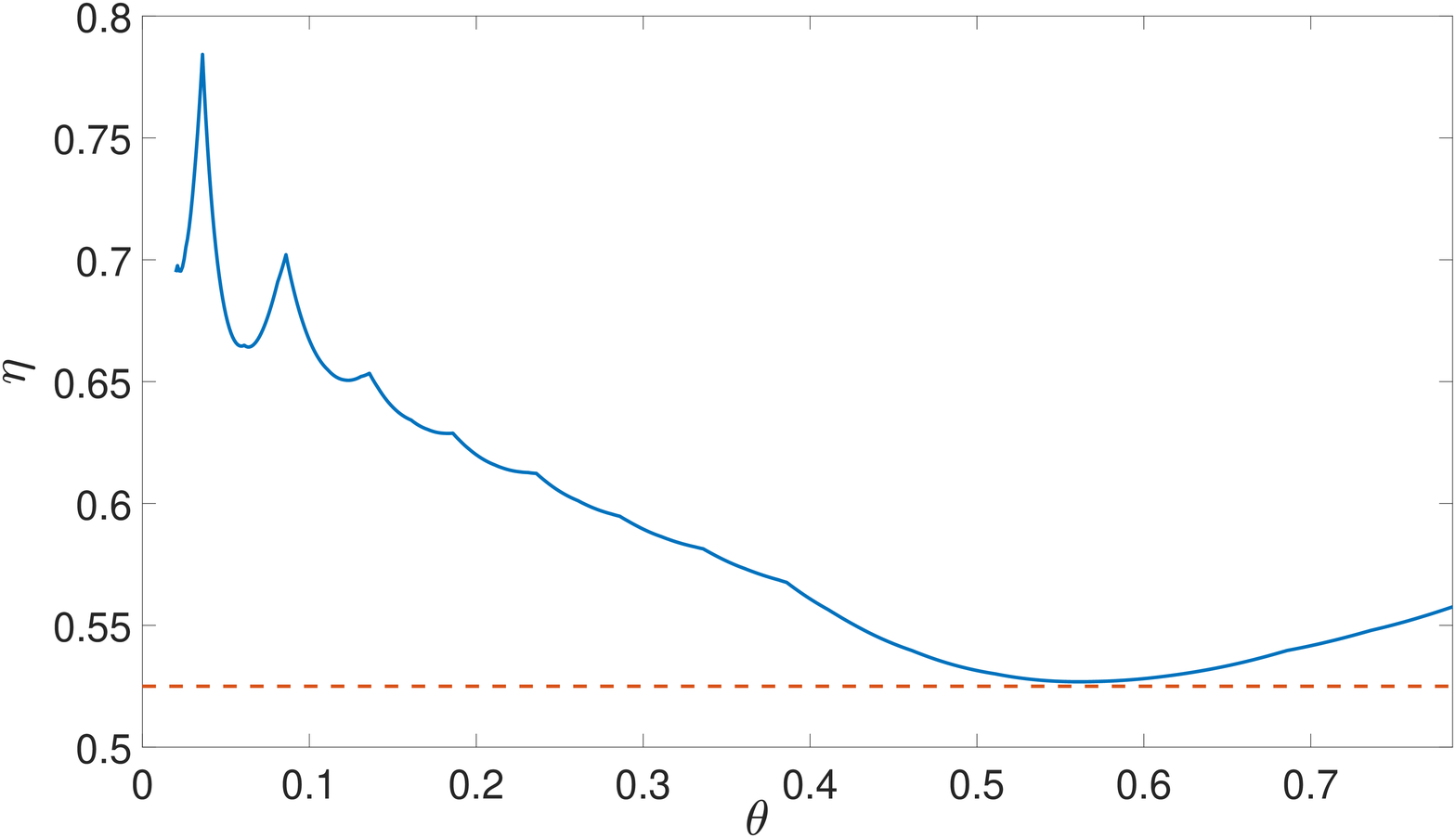}
\caption{The blue curve, delimits the parameter region where the state $\rho_\theta^{\eta}$ admits a local model for projective measurements on Alice's side and POVMs on Bob's side. Therefore, all states $\rho_\theta^{\eta^*}$ are local, with $\eta^* = 0.525$ and for all $\theta \in [ 0 , \pi/4]$.
}
\end{center}
\end{figure}

\subsection{Step 1: constructing a net of local states}

The efficiency of the algorithm of Refs \cite{hirsch15,cavalcanti15} relies on using a finite set of measurements that represents a good approximation of the (continuous) set of measurements  which the model should work for. In particular, when considering qubit projective measurements, a geometrical approach based on the Bloch sphere can be applied efficiently, see e.g. \cite{hirsch16}. For qubit POVMs however, the situation is much more involved as the Bloch sphere does no longer provide a complete description. 
 Indeed, any extremal qubit POVM has at most four outcomes \cite{dariano05}. Thus, it is described by a set of four POVM elements, which can be parametrized by 12 real numbers (one needs to characterize three $2 \times 2$ positive semi-definite matrices; the fourth one being determined by normalisation). On the other hand, qubit projective measurements require only two real parameters to be characterised

This problem can however be addressed. Here we use a slightly improved version of Protocol 2 of \cite{hirsch16}, given in details Appendix A. In this version, the algorithm can take advantage of certain auxiliary states (initially inputted in the code) that are known to admit a local model. 

We thus first construct a list of useful auxiliary states that admit a LHV model with the desired property, namely considering all projective measurements for Alice and all POVMs for Bob. 

\emph{Class 1.} We start by constructing LHV models for states $\rho_{\theta}^{\eta}$ for projective measurements (on both sides) using the method of \cite{hirsch16}. 

More precisely, Ref. \cite{hirsch16} presented an efficient implementation of the algorithm of Refs \cite{hirsch15,cavalcanti15}, combined with another numerical method \cite{brierley16}, tailored for the case of the two-qubit Werner state, i.e. $\rho_{\pi/4}^{\eta}$. A LHV model was presented for $\eta \simeq 0.68$. Here we used a similar implementation for states $\rho_{\theta}^{\eta}$, with $\theta = 0.65$ and $\theta = 0.6$. In these two cases, we find that the state  $\rho_{\theta}^{\eta}$ admits a LHV for projective measurements (on both sides) for $\eta = 0.69$. Note that this construction, while being based on a numerical heuristic search, can in principle be made fully analytical using the method described in Appendix C of Ref. \cite{hirsch16}.

From this, we now construct entangled states admitting a local model for projective measurements for Alice and POVMs on Bob's side. More precisely we make use of Lemma 2 of Ref. \cite{hirsch16} (see also \cite{oszmaniec16}). This tells us that a noisy qubit POVM with elements of the form 

\ba 
N_{b|y}^{\mu}  = \mu  N_{b|y} + (1-\mu) \frac{\openone}{4}
\ea
can always be simulated from projective measurements (i.e. for any qubit POVM $N_{b|y}$) when $\mu \leq \sqrt{2/3}$. Specifically, the statistics of such a noisy POVM can be obtained from a model for projective measurements followed by classical post-processing \cite{guerini17}. 

This implies that states of the form
\begin{align}  \label{class1}
\xi_{\theta}^{\eta} = \sqrt{ \frac{2}{3} } \rho_{\theta}^{\eta} + (1-\sqrt{\frac{2}{3}}) \xi_A \otimes \frac{\openone}{2}
\end{align}
where $\xi_A = \tr_B(\rho_{\theta}^{\eta})$, now admit a LHV models for POVMs on Bob's side, as desired. The states $\xi_{\theta}^{\eta} $ are thus added to the list.

\emph{Class 2.} We start from the class of states
\begin{align}
\xi_\theta =  \alpha \ket{\phi_\theta} \bra{\phi_\theta} + (1-\alpha)  \frac{\openone}{2} \otimes \rho_B
\end{align}
where the parameter $\alpha$ is given by 
\begin{align} 
\label{condition} \cos^2(2\theta)=\frac{2\alpha-1}{(2-\alpha) \alpha^3}.
\end{align} 
These states admit a local model for POVMs on Alice and projective measurements for Bob \cite{bowles15b}. 
Note that states $\xi_\theta $ are of the form \eqref{Maf}. We now make use 
of the extension technique of Ref. \cite{hirsch13}; specifically, from Protocol 2 we have that states of the form 
\begin{align} \label{rho_th}
\gamma_{\theta} =  \frac{1}{2} ( \xi_\theta  + \xi_A \otimes \ket{0}\bra{0} )
\end{align}
where $\xi_A = \tr_B (\xi_\theta)$, now admit a LHV model for POVMs on both sides. The states $\gamma_{\theta}$ are added to the list of auxiliary states.

\emph{Class 3.} Finally, we use the states  
\begin{align}
\beta_\theta =  \alpha \ket{\phi_\theta} \bra{\phi_\theta} + (1-\alpha) \rho_A \otimes \frac{\openone}{2}
\end{align}
with $\rho_A = \tr_B(\ket{\phi_\theta} \bra{\phi_\theta} )$, and where the parameter $\alpha$ is given by equation \eqref{condition}.
Note that states $\beta_\theta $ are of the form \eqref{Maf} up to a permutation of Alice and Bob.

This completes our list of auxiliary states. Next, we sample the interval $\theta \in [0,\pi/4]$ with $16$ equally spaced values, which we denote $\vec{\theta} = \pi/4:-0.05:0$. For each value we run the algorithm, including the list of auxiliary states. We make use of $46$ projective measurements for Alice and $6$ projective measurements for Bob (details in Appendix A). For a given value of $\theta_k$, the algorithm returns a value $\eta_k$ such that $\rho_{\theta_k}^{\eta_k}$ is local.

We then run the algorithm again, now adding the previously obtained set of states $\rho_{\theta_k}^{\eta_k}$. Here construct a finer net $\vec{\theta} = \pi/4:-0.005:0$. We use $16$ projective measurements for Alice and $6$ projective measurements for Bob. 

Finally, we do a last iteration by taking only $6$ measurements on both sides and $\vec{\theta} = \pi/4:-0.001:0$. 

Note that these iterations are done for practical convenience, as they reduce computation time. In principle, one could use directly a fine grid at the first iteration.

In practice, this is limited by computer precision and may lead to
some false results in some pathological cases.
Again, this issue can in principle be circumvented by using similar techniques to those described in Appendix C of Ref. \cite{hirsch16} and working interval arithmetic see Ref. \cite{vsdp}, thus making the construction fully analytical.
Typically, implementing this procedure will result in a slight decrease of the final visibility, in most of the cases, of the order of the SDP precision ($\approx 10^{-8}$ ), which will essentially not change the result.

\subsection{Continuity arguments}

Now that we have obtained a fine net of local states, we must extend the result to the continuous interval. 

We first consider the regime of small $\theta$. We use the explicit model for projective measurements of $\rho_{\theta}^{\eta}$ given in Ref. \cite{bowles15b}. The model works for $\eta$ given by condition \eqref{condition}. To take POVMs into account we use again the method of Ref. \cite{hirsch13}. Specifically, we now apply Protocol 2 to the state $\rho_{\theta}^{\eta}$ such that condition \eqref{condition} is fulfilled. We have thus again a class of states which admit a LHV model for all projective measurements on Alice's side and all POVMs for Bob. 

The last step consists in showing that $\rho_{\theta}^{\eta^*}$ can be written as a convex combination of this class of states and a separable two-qubit state, for all $\theta \in [0,\theta_0]$, for some $\theta_0 > 0$. The proof is given in Appendix B.

Second, we move to the regime $\theta \in [\theta_0, \pi/4]$. We make use of the following result: 

{\bf Lemma 1.} Consider that the state $\rho_{\theta}^{\eta}$ admits a LHV model for sets of (possibly infinitely many) measurements $\mathcal{M}_A$ for Alice and $\mathcal{N}_B$ for Bob. Then the state $\rho_{\theta'}^{\eta'}$, with $\theta' \geq \theta$, also admits a LHV model (for the same sets $\mathcal{M}_A$, $\mathcal{N}_B$) as long as
\begin{align} \label{lemma1}
\tan(\theta') \frac{\eta'}{(1+\eta')} \leq \tan(\theta) \frac{\eta}{(1+\eta)}. 
\end{align}

This lemma is proven in Appendix B of \cite{hirsch16}. It allows us to extend a model for the state $\rho_{\theta}^{\eta}$ to the continuous interval $[\theta,\theta']$, with visibility $\eta'$ given by \eqref{lemma1}. That is, we ensured that the state $\rho_{\theta'}^{\eta'}$ admits a model in the range $[\theta,\theta']$. 

We thus use Lemma 1 on each interval $[\theta_k,\theta_{k+1}]$ of our net. This gives us some value $\eta_k$ such that $\rho_{\theta_k}^{\eta_k}$ admits a model for all $\theta \in [\theta_k,\theta_{k+1}]$. Finally, we conclude that $\rho_{\theta_k}^{\eta^*}$ admits a LHV model for all $\theta \in [0, \pi/4]$, where $\eta^* = \min_k (\eta_k) = 0.525$.

\section{Discussion}

We proved that measurement incompatibility does not imply Bell nonlocality in general. Specifically, we have shown that a given set of non-jointly-measurable qubit POVMs can never lead to Bell inequality violation, as it admits a LHV model. Our construction is general, as it considers any possible shared entangled state, and any possible measurements performed by the second observer. We note that a similar result was recently proven by Bene and V\'ertesi \cite{erika17}.

An interesting question is to find the minimal setting (in terms of number of POVMs or number of measurement outcomes) in which this results holds. While, our construction involves a set of infinitely many qubit POVMs, one can easily adapt it to construct examples with finitely many POVMs. For instance, in reference  \cite{bavaresco17} the authors presented a set of seven qubit measurements of the form \eqref{POVM} (with vectors $\hat{x}$ chosen rather uniformly on the Bloch sphere) that is incompatible for visibility $\eta=0.524<\min_k (\eta_k) = 0.525$,  being then an explicit example of a set of finite incompatible Bell local measurements. We note that in the simplest case, namely two binary POVMs, measurement incompatibility does imply Bell nonlocality \cite{wolf09}; the CHSH Bell inequality being enough here. Moving away from this simplest case, things change. In particular, for sets of three qubit measurements, Bene and V\'ertesi \cite{erika17} could recently prove that measurement incompatibility does not imply Bell nonlocality. To complete these results, it would be interesting to study the case of two d-outcome POVMs.

Finally, it would be interesting to see whether there exists a natural notion of measurement incompatibility (stronger than joint measurability) which would always lead to Bell nonlocality.

\emph{Acknowledgements.} We thank Joe Bowles and Tam\'as V\'ertesi for discussions. We acknowledge financial support from the Swiss National Science Foundation (Starting grant DIAQ and QSIT) and  Japan Society for the Promotion of Science (JSPS) by KAKENHI grant No. 16F16769.

\appendix

\section{Algorithm for LHV models}

As stated in the main text we used the algorithmic construction of \cite{hirsch16} to find LHV models for states $\rho_\theta^{\eta}$. More precisely we note that for a fixed $\theta=\theta_f$ the state is linear with respect to $\eta$ and we can thus use Protocol 2 of \cite{hirsch16} to find an $\eta=\eta_f$ such that $\rho_{\theta_f}^{\eta_f}$ admits a LHV model. 

We have run a slightly improved version of Protocol 2, which requires to choose finite sets of POVMs $ \{M_{a|x} \}$ and $ \{N_{b|y} \}$ (with associated parameters $\nu$, $\mu$), density matrices $\xi_A$ and $\xi_B$ and auxiliary states $\rho_k$. We then run the following SDP:

{\bf Protocol 2.} (improved version)
\ba \text{find  } & &  q^* = \max  q  \\
\text{s.t. }   & & Tr(M _{a|x} \otimes N_{b|y} \chi) = \sum_\lambda p_\lambda D_\lambda(ab|xy)  \,\,  \forall a,b,x,y   \nonumber  \\ \nonumber    
& & p_\lambda \geq 0 \,\,  \forall \lambda  \\ \nonumber
& & \chi^{\nu,\mu} = \nu \mu \chi + \nu (1-\mu) \chi_A \otimes \xi_B  + \mu (1-\nu) \xi_A \otimes \chi_B +   \\ & & \nonumber
\quad  (1-\nu) (1-\mu) Tr(\chi) \xi_A \otimes \xi_B  \\ \nonumber
& & \rho_\theta^{q}  - \chi^{\nu,\mu} - \sum_k \beta_k \rho_k \geq 0 \\ \nonumber
& & (\rho_\theta^{q}  - \chi^{\nu,\mu} - \sum_k \beta_k \rho_k)^{T_B} \geq 0 \\ \nonumber
& & Tr(\chi) \geq 0 \\ \nonumber
& & \beta_k  \geq 0 \,\,  \forall k \\ \nonumber
\ea
where $^{T_B}$ stands for the partial transposition on Bob's side and the optimization variable are (i) positive coefficients $p_\lambda$ and $\beta_k$ and (ii) a $d \times d$ hermitian matrix $\chi$. Given $m_A$ ($m_B$) dichotomic measurements, one has $N = 2^{m_A} 2^{m_B}$ local deterministic strategies $D_\lambda(ab|xy)$, and $\lambda = 1,...,N $.

For the answer to hold (i.e. ensuring that $\rho_{\theta_f}^{\eta}$ admits a LHV model) the parameters $\nu, \mu$ must be smaller or equal to the `shrinking factors' of the set of all POVMs with respect to the finite set  $ \{M_{a|x} \}$ (and given state $\xi_A$), respectively  $ \{N_{b|y} \}$ (and given state $\xi_B$). That is, the largest $\nu = \nu^*$  such that any shrunk POVM $\{M^{\eta}_a\}$ with elements defined by
\begin{align}
M_{a}^\eta = \eta M_{a} + (1-\eta) \Tr[\xi_A M_a] \openone
\end{align}
can be written as a convex combination of the elements of $ \{M_{a|x} \}$, i.e. $M_{a}^\eta = \sum_{x} p_x M_{a|x} $ ($\forall a$) with $\sum p_x=1$ and $p_x\geq 0$ (and similarly for measurements $\{N_{b|y} \}$ and given state $\xi_B$) . The exact value $\nu^*$ is in general hard to evaluate, but Ref. \cite{hirsch16} gives a general procedure to obtain arbitrary good lower bounds on $\nu^*$, which is therefore enough for us to make sure that $\nu \leq \nu^*$.

\subsection{Choice of the algorithm parameters}

We first took $16$ equally spaced values in $[0,\pi/4]$, which we denote $\vec{\theta} = \pi/4:-0.05:0$. For each $\theta_k$ we used the same parameters on Alice's side, but different ones on Bob's side.

\subsubsection{Alice's side}
Let us start with the parameters on Alice's side. We first set $\xi_A = \frac{\openone }{2}$. The finite set $ \{M_{a|x} \}$ must consist of projective measurements, which can be represented by vectors on the Bloch sphere. We picked $46$ of them, corresponding to the following procedure: we start from an icosahedron (\textit{i.e.} $12$ pairwise opposite vertices, representing $6$ measurements), we then add 10 new measurements (i.e. 20 vertices) corresponding to the geometrical dual of the icosahedron. We thus get a new set with 16 measurements (32 vertices). We do another iteration (i.e. adding vertices of the dual) and obtain a final polyhedron with 92 vertices, and a corresponding shrinking factor of $\eta_3 \approx 0.971$.

Note that the number of deterministic strategies $N = 2^{m_A} 2^{m_B}$ becomes huge for this case, in which $m_A = 46$. However, we used the technique of \cite{hirsch16} (Appendix D), which consists of restricting ourselves to deterministic strategies compatible with the sign response function used in Werner's model \cite{werner89}, leading to only $1772$ deterministic strategies out of the $2^{46}$ available to Alice.

\subsubsection{Bob's side}

Motivated by the form of the states $\rho_\theta^{\eta}$ we adapt the algorithm parameters of Bob in function of $\theta$. More precisely for each $\theta_k$ we set $\xi_B = \rho_B$ and pick the $6$ projective measurements which maximizes the shrinking factor $\mu$ for this choice of $\xi_B$. This amounts in $16$ different polyhedrons, associated to the $16$ values of $\vec{\theta}$. However, we need to ensure the existence of a model for all POVMs for Bob. In order to do so we consider all relabellings of $ \{ P_+ , P_- , 0 , 0 \}$ for $P_+$ being a projector onto a vertex of the polyhedron in the Bloch sphere and $P_-$ onto the opposite direction (all $16$ polyhedrons have pairwise opposite vertices). In addition we consider the four relabellings of the trivial measurement $ \{ \openone  , 0 , 0 , 0 \}$, which comes for free as it cannot help to violate any Bell inequality and consequently does not even need to be inputed in Protocol 2. The set thus have 76 elements, but we need to take into account only 6 of them when running the Protocol, corresponding to the vertices in the upper half sphere of the polyhedron of interest.

We thus computed lower bounds on the shrinking factors for different $\xi_B$. All bounds for $\mu^*$ in function of $\theta$ are given in Tables 1 and 2.

\begin{table}[t!] \label{table1}
\begin{tabular}{| c || c|c|c|c|c|c|c|c|c|c|}
  \hline
 $\theta$  & 0.7854 & 0.7354  &  0.6854   & 0.6354  &  0.5854  &  0.5354  &  0.4854   & 0.4354     \\
 \hline
 $\mu$ &  0.6737  &   0.6728 &    0.6715 &    0.6698 &    0.6676 &    0.6647 &    0.6584   &  0.6500    \\
 \hline
\end{tabular} \caption{Lower bounds on shrinking factors of the set of qubit POVMs with respect to different finite set of POVMs coming from polyhedrons in the Bloch sphere and with $\xi_A = \cos(\theta)^2 \ketbra{0}{0} + \sin(\theta)^2 \ketbra{1}{1}	$}
 \end{table}

\begin{table}[t!] \label{table2}
\begin{tabular}{| c || c|c|c|c|c|c|c|c|c|c|}
  \hline
 $\theta$  & 0.3854 &   0.3354   & 0.2854  &  0.2354  &  0.1854  &  0.1354  &  0.0854 &  0.0354  \\
 \hline
 $\mu$  & 0.6385  &  0.6202   &  0.6038  &   0.5601   &  0.5040   &  0.4391   &  0.3487 & 0.2143 \\
 \hline
\end{tabular} \caption{Lower bounds on shrinking factors of the set of qubit POVMs with respect to different finite set of POVMs coming from polyhedrons in the Bloch sphere and with $\xi_A = \cos(\theta)^2 \ketbra{0}{0} + \sin(\theta)^2 \ketbra{1}{1}	$}
 \end{table}

\begin{table}[t!] \label{table3}
\begin{tabular}{| c || c|c|c|c|c|c|c|c|c|c|}
  \hline
 $\theta$  & 0.7854 & 0.7354  &  0.6854   & 0.6354  &  0.5854  &  0.5354  &  0.4854   & 0.4354     \\
 \hline
 $\eta$ &  0.5577  &   0.5478  &   0.5396  &   0.5231   &  0.5124  &   0.5127  &   0.5215   &  0.5400    \\
 \hline
\end{tabular} \caption{Outputs of Protocol 2 for each $\theta_k$ in step 1}
 \end{table}

\begin{table}[t!] \label{table4}
\begin{tabular}{| c || c|c|c|c|c|c|c|c|c|c|}
  \hline
 $\theta$  & 0.3854 &   0.3354   & 0.2854  &  0.2354  &  0.1854  &  0.1354  &  0.0854 &  0.0354  \\
 \hline
 $\eta$ & 0.5679 &    0.5815 &    0.5953 &    0.6127  &   0.6291  &   0.6545 &    0.7021 & 0.7902\\
 \hline
\end{tabular} \caption{Outputs of Protocol 2 for each $\theta_k$ in step 1}
 \end{table}

\subsubsection{Final implementation}

The first iteration consists of $16$ uses of Protocol 2 with the parameters given above. The obtained values of $\eta$ are given in Table 3 and 4. We then add the states $\rho_{\theta_k}^{\eta_k}$ to the list of auxiliary states $\rho_k$. We run Protocol 2 again for $\vec{\theta} = \pi/4:-0.005:0$, using only $16$ measurements for Alice, represented in the Bloch sphere by the union of an icosahedron and its dual. For Bob, we pick the closest $\theta_k$ to $\theta$ and use the corresponding $\xi_B= \rho_B$ alongside with its ''optimal'' polyhedron (with known bound on the shrinking factor of POVMs $\mu$). Finally we repeat the same procedure for $\vec{\theta} = \pi/4:-0.001:0$, using  $6$ measurements for Alice, represented in the Bloch sphere by an icosahedron. This gives us the final plot of Figure 1.

%

\section{LHV model for small $\theta$ }

Here we give the proof that for all $\theta \in [0,0.02]$ the state $\rho_\theta^{\eta*}$ can be written as a convex combination of a separable state and $\chi_\theta^{\beta} =  \frac{1}{2} ( \rho_\theta^{\beta}  + \rho_A \otimes \ket{0}\bra{0} )$, where $\rho_A = \Tr_B(\rho_\theta^{\beta})$ and $\beta$ and $\theta$ are linked by 

\ba \label{condition2}
\cos(2\theta)^2 \geq \frac{2\beta-1}{(2-\beta) \beta^3}
\ea

ensuring thus that $\chi_\theta^{\beta}$ admits a LHV models for all POVMs on both sides, as explained in the main text.

We want:
\begin{align}
\rho_\theta^{\eta*} = q \chi_\theta^{\beta} + (1-q) S
\end{align}
where $S$ is a separable state. Inverting this relation we get: 
\begin{align}
(1-q) S = \rho_\theta^{\eta*}- q \chi_\theta^{\beta}.
\end{align}
Let us define the matrix $A = 4 (1-q) S$. The non-zero elements of $A$ are given by:
\begin{align} 
&A(1,1)=2\cos^2\theta (1+\eta^*) - q (\cos^2\theta (1+\beta) \nonumber +\\
& \quad \quad 2\cos^2\theta \beta +(1-\beta))  \nonumber \\ 
&A(2,2)=2 \sin^2\theta (1-\eta^*) - q \sin^2\theta (1-\beta) \nonumber \\ 
&A(3,3)=2\cos^2\theta (1-\eta^*) - q (\cos^2\theta (1-\beta) +  \nonumber \\
& \quad \quad  2 \sin^2\theta \beta + (1-\beta)) \nonumber \\ 
&A(4,4)=2 \sin^2\theta (1+\eta^*) - q \sin^2\theta (1+\beta)  \nonumber \\
&A(1,4)=S(4,1)=4 \eta^* \cos\theta \sin\theta -  2 q \beta \cos\theta \sin\theta  \nonumber	.
\end{align}

Now let us set $\theta=0.02$, $q = 0.682$ and $\beta=0.911$ (and recall that $\eta^* = 0.525$). First one can check that relation given by Eq. 16 holds, ensuring that $\chi_{0.02}^{0.911}$ admits a LHV models for all POVMs (on both sides). For those values one can also check that $A$ is positive and has a positive partial transpose. Let us now consider $0 < \theta < 0.02$ (keeping the same values for the other parameters): relation \eqref{condition2} still holds has the right-hand term increases while the left-hand one remains constant. We have the following conditions for $A$ to be positive with a positive partial transpose: 

\begin{align} 
&A(k,k) \geq 0 \; \; k=1,2,3,4 \label{cond1} \\
&A(1,1)\cdot A(4,4) - A(1,4)^2 \geq 0 \label{cond2} \\ 
&A(2,2)\cdot A(3,3) - A(1,4)^2 \geq 0 \label{cond3} 
\end{align}

which, if true for some $\theta_0 \in [0,\pi/4]$, are true for $0 < \theta \leq \theta_0$. Indeed, there are only two types of equations, first the cases $k=2$ and $k=4$ of \eqref{cond1}, which are of the form $P \cdot \sin(\theta)^2$, whose sign does not depend on $\theta$, and the other cases, which are of the form $A \cos(\theta)^2  - B \sin(\theta)^2 - C \geq 0$, where $A,B,C \geq 0$. If all fulfilled for $\theta=0.02$, those conditions are therefore also fulfilled for $\theta \in [0,0.02]$, as in this regime and for any $A,B,C \geq 0$ one has $A \cos(\theta)^2  - B \sin(\theta)^2 - C \geq A \cos(\theta=0.02)^2  - B \sin(\theta=0.02)^2 - C \geq 0$ . 

Finally, taking $S=\frac{A}{4(1-q)}$ together with the above parameters we conclude that $\rho_\theta^{\eta*} = q \chi_\theta^{\beta} + (1-q) S$ (for any $\theta \in [0,0.02]$) with $S$ a (valid) separable state (via the partial transposition criterion of \cite{horodecki_ppt}).

\bibliographystyle{linksen}  
\bibliography{mtqbib}        

\end{document}